\documentclass[aps,prb,twocolumn,superscriptaddress,floatfix]{revtex4}
\usepackage{graphicx,color}
\usepackage{dcolumn}
\hfuzz=\maxdimen
\begin{document}


\title{Growth and Characterization of Millimeter-sized Single Crystals of CaFeAsF}

\author{Yonghui Ma}
\affiliation{School of Physical
Science and Technology, ShanghaiTech University, Shanghai 201210,
China} \affiliation{State Key Laboratory of Functional Materials for
Informatics and Shanghai Center for Superconductivity, Shanghai
Institute of Microsystem and Information Technology, Chinese Academy
of Sciences, Shanghai 200050, China}

\author{Hui Zhang}
\affiliation{CAS Key Laboratory of Materials for Energy Conversion,
Shanghai Institute of Ceramics, Chinese Academy of Sciences,
Shanghai 200050, China}
\author{Bo Gao}
\affiliation{State
Key Laboratory of Functional Materials for Informatics and Shanghai
Center for Superconductivity, Shanghai Institute of Microsystem and
Information Technology, Chinese Academy of Sciences, Shanghai
200050, China}
\author{Kangkang Hu}\affiliation{State Key Laboratory of Functional Materials for
Informatics and Shanghai Center for Superconductivity, Shanghai
Institute of Microsystem and Information Technology, Chinese Academy
of Sciences, Shanghai 200050, China} \affiliation{College of
Sciences, Shanghai University, Shanghai 200444, China}
\author{Qiucheng Ji}
\affiliation{State Key Laboratory of Functional Materials for
Informatics and Shanghai Center for Superconductivity, Shanghai
Institute of Microsystem and Information Technology, Chinese Academy
of Sciences, Shanghai 200050, China}
\author{Gang Mu}
\email[]{mugang@mail.sim.ac.cn}
\affiliation{State Key Laboratory of Functional Materials for Informatics and Shanghai Center for Superconductivity, Shanghai Institute of Microsystem and Information Technology, Chinese Academy of Sciences, Shanghai 200050, China}
\author{Fuqiang Huang}
\affiliation{CAS Key Laboratory of Materials for Energy Conversion,
Shanghai Institute of Ceramics, Chinese Academy of Sciences,
Shanghai 200050, China}
\author{Xiaoming Xie}
\affiliation{School of Physical Science and Technology, ShanghaiTech
University, Shanghai 201210, China}\affiliation{State Key Laboratory
of Functional Materials for Informatics and Shanghai Center for
Superconductivity, Shanghai Institute of Microsystem and Information
Technology, Chinese Academy of Sciences, Shanghai 200050, China}

\begin{abstract}
High-quality and sizable single crystals are crucial for studying
the intrinsic properties of unconventional superconductors, which
are lacking in the 1111 phase of the Fe-based superconductors. Here
we report the successful growth of CaFeAsF single crystals with the
sizes of 1-2 mm using the self-flux method. Owning to the
availability of the high-quality single crystals, the structure and
transport properties were investigated with a high reliability. The
structure was refined by using the single-crystal x-ray diffraction
data, which confirms the reports earlier on the basis of powder
data. A clear anomaly associated with the structural transition was
observed at 121 K from the resistivity, magnetoresistance, and
magnetic susceptibility measurements. Another kink-feature at 110 K,
most likely an indication of the antiferromagnetic transition, was
also detected in the resistivity data. Our results supply a basis to
propel the physical investigations on the 1111 phase of the Fe-based
superconductors.

\end{abstract}

\maketitle

\section{introduction}
The F-doped LnFeAsO (Ln = rare-earth elements), which has been
abbreviated as 1111 phase, is the first reported family with the
highest critical transition temperature $T_c$ in bulk in the
Fe-based superconductors (FeSCs).~\cite{Hosono} However,
investigations on the physical properties of 1111-type FeSCs are
restricted remarkably, compared with the 122 phase and 11 phase, due
to the difficulties in obtaining sizable single crystals. As we
know, it is essential to have high-quality single crystals when
carrying out many experiments, including the measurements of
electrical transport, inelastic neutron diffraction, angle resolved
photoemission spectroscopy, and so on. During the past several
years, many efforts have been made to improve the quality and size
of the single crystals. NaCl and KCl were first used as the flux and
small single crystals with sizes of 20-70 $\mu$m can be
obtained.~\cite{NaCl} Then more attempts, including the
high-pressure method and the NaAs-flux method,~\cite{NaAs,HP} were
made to further improve the growth processes. Up to now, the two
goals, sizable and high-quality, are still not achieved commendably.
Recently, single crystals with the size of several millimeters were
reported to be accessible in F-vacant and Na-doped
CaFeAsF,~\cite{F-vacant,Na-doped} which is another type of 1111
phase without oxygen,~\cite{SrFeAsF1,SrFeAsF2} possibly due to the
decrease of melting point in this fluorine-based system. As we know,
a rather high $T_c$ above 50 K can also be achieved by doping in
this fluorine-based 1111
system.~\cite{CaFeAsF-Nd,CaFeAsF-Co,SrFeAsF-Sm,SrFeAsF-La} More
important information can be obtained owning to the availability of
the sizable single crystals. To our knowledge, the investigations on
the single crystals of the parent phase CaFeAsF are still lacking.

Here we present the growth, structure, and transport measurements of
the high-quality CaFeAsF single crystals with the sizes of 1-2 mm.
The single crystals were grown by the self-flux method. The
structure details were obtained from the refinement of the
single-crystal x-ray diffraction data. The structural transition at
121 K was confirmed by the resistivity, magnetoresistance, and
magnetic susceptibility measurements. A feature coming from the
antiferromagnetic transition was also observed in the resistivity
data.

\section{Experimental details}
High quality CaFeAsF single crystals
were grown using the self-flux method with CaAs as the flux. First,
the starting materials Ca granules (purity 99. 5\%, Alfa Aesar) and
As grains (purity 99.995\%, Alfa Aesar) were mixed in 1: 1 ratio.
Then the mixture was sealed in an evacuated quartz tube and followed
by a heating process at 700$^\circ$C for 10 h to get the CaAs
precursor. CaAs, FeF2 powder (purity 99\%, Alfa Aesar) and Fe powder
(purity 99+\%, Alfa Aesar) were mixed together in the stoichiometric
ratio 10: 1: 1, and the mixture were placed in a crucible. Finally,
the crucible was sealed in a quartz tube with vacuum. All the
weighing and mixing procedures were carried out in a glove box with
a protective argon atmosphere. The quartz tube was heated at
950$^\circ$C for 40 hours firstly, and then it was heated up to
1230$^\circ$C and stay for 20 hours. Finally it was cooled down to
900$^\circ$C at a rate of 2$^\circ$C /h and followed by a quick
cooling down to room temperature.

The microstructure was examined by the scanning electron microscopy
(SEM, Zeiss Supra55). The composition of the single crystals was
checked and determined by the energy dispersive x-ray spectroscopy
(EDS) measurements on an Oxford Instruments. The crystals were first
checked using a DX-2700 type powder x-ray diffractometer. The
detailed structure was characterized and analyzed by the
single-crystal x-ray diffraction measurements on a Bruker D8 Focus
diffractometer equipped with the graphite-monochromatized Mo
$K_\alpha$ radiation. The magnetic susceptibility measurement was
carried out on the magnetic property measurement system (Quantum
Design, MPMS 3). The electrical resistance and magnetoresistance
(MR) were measured using a four-probe technique on the physical
property measurement system (Quantum Design, PPMS) with magnetic
field up to 9 T. For the MR measurements, the magnetic field was
oriented parallel to the c axis of the samples and the data were
measured for both positive and negative field orientations to
eliminate the effect of the Hall signals.

\section{Results and discussions}
A typical dimension of the single crystals is
1.2$\times$1.0$\times$0.1 mm$^3$. The morphology was examined by the
scanning electron microscopy. An SEM picture for the CaFeAsF single
crystal can be seen in Fig. 1(a), which shows the flat surface and
some terrace-like features. An enlarged view of this picture can be
seen in Fig. 1(b). The composition of the crystals was characterized
by energy-dispersive x-ray spectroscopy (EDS) measurements. We
measured the EDS at different positions of the sample. Here we show
a typical result in Fig. 1(c) and Table I, which revealed that the
ratio of Ca: Fe: As is close to the stoichiometric ratio. The
content of the light element F is difficult to determine precisely
based on EDS measurements. The structure of the crystals was first
check by a powder x-ray diffractometer, where the x-ray was incident
on the ab-plane of the crystal. The diffraction pattern is shown in
Fig. 2. All the diffraction peaks can be indexed to the tetragonal
ZrCuSiAs-type structure (see the inset of Fig. 2). Only sharp peaks
along (00l) orientation can be observed, suggesting a high c-axis
orientation. The full width at half maximum (FWHM) of the
diffraction peaks is only about 0.10$^\circ$ after deducting the
$K_{\alpha2}$ contribution, indicating a rather fine crystalline
quality. The c-axis lattice constant was obtained to be 8.584 {\AA}
by analyzing the diffraction data.

\begin{figure}
\includegraphics[width=9.5 cm]{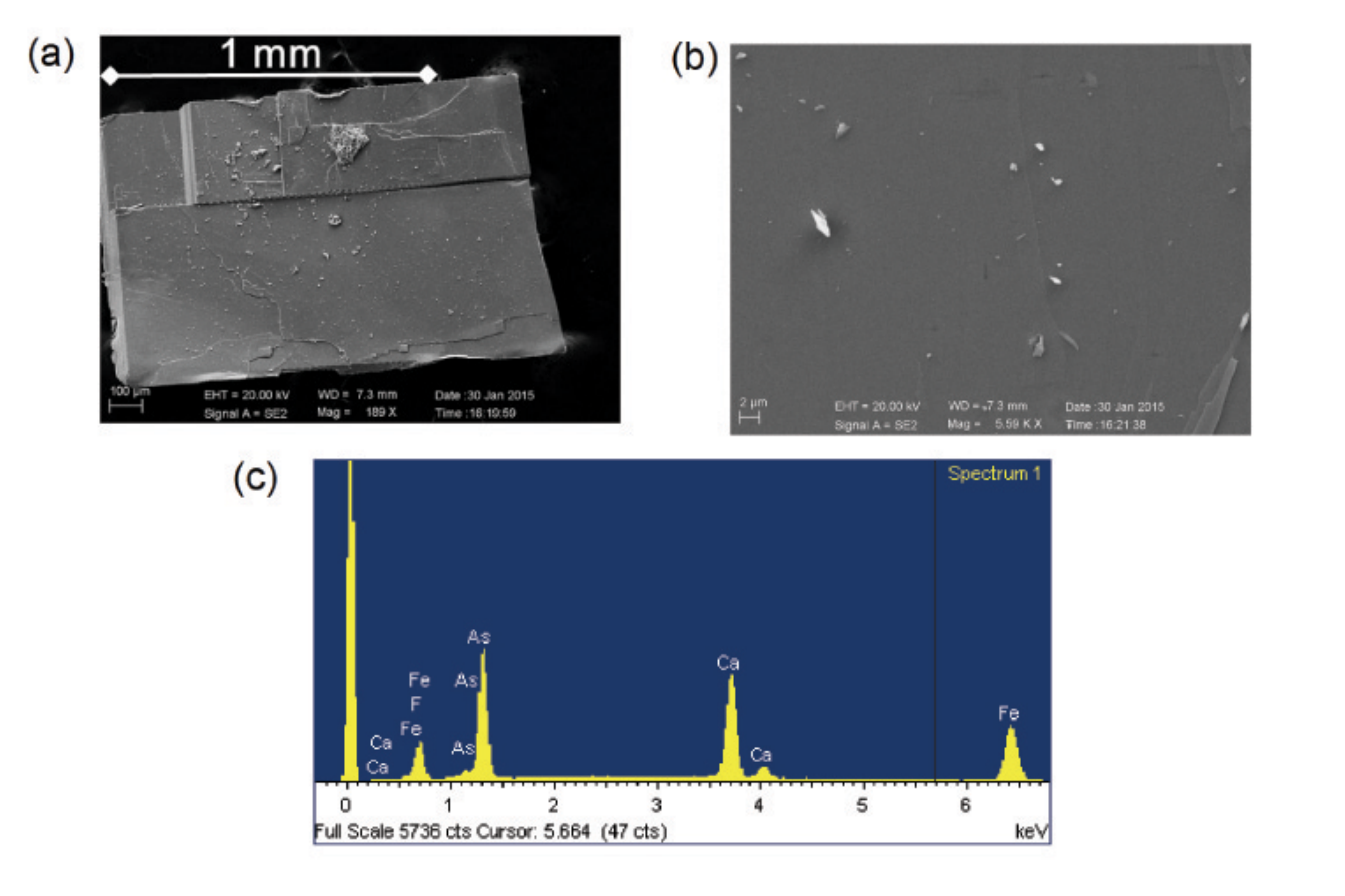}
\caption {(color online) (a) An SEM picture of a CaFeAsF crystal
with the lateral size larger than 1 mm. (b) The enlarged view of the
SEM picture. (c) The EDS microanalysis spectrum taken on one
crystal.} \label{fig1}
\end{figure}

\begin{table}
\centering \caption{Compositions of the crystal characterized by EDS
measurements.}
\begin{tabular}
{ccccccc}\hline \hline
Element &       Weight (\%)   &  Atomic (\%)  \\
\hline
F          & 14.90   & 34.52    \\
Ca    & 19.58   & 21.50   \\
Fe         & 27.31   & 21.52   \\
As           & 38.22   & 22.46   \\
 \hline \hline
\end{tabular}
\label{tab.1}
\end{table}

\begin{figure}
\includegraphics[width=8.8cm]{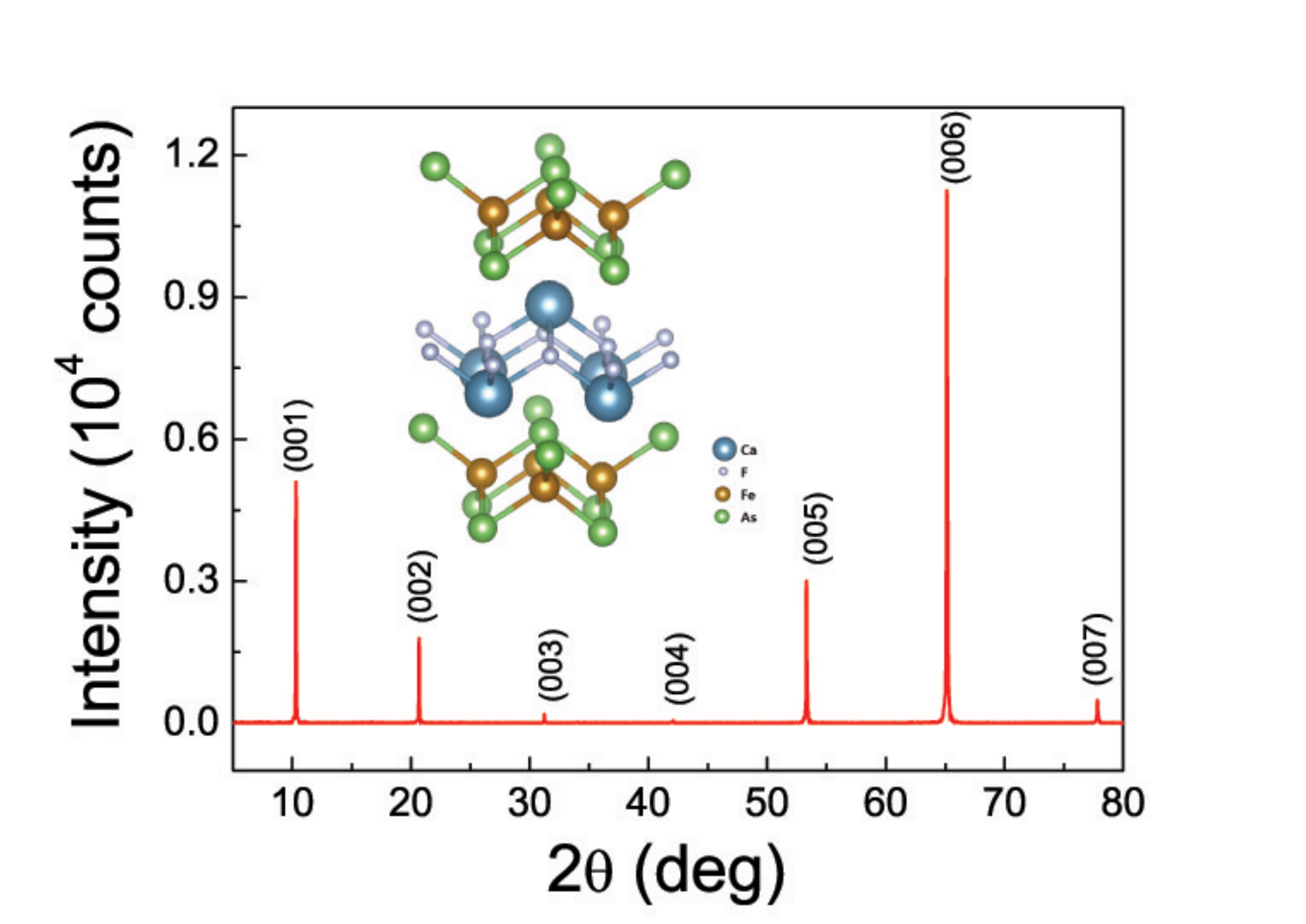}
\caption {(color online) X-ray diffraction pattern measured on the
CaFeAsF single crystal with the x-ray incident on the $ab$-plane.
The inset is the schematic of the crystal structure of CaFeAsF.}
\label{fig2}
\end{figure}

\begin{table}
\centering \caption{Parameters for the data collection and structure
refinement of CaFeAsF.}
\begin{tabular}{ll}
    \hline
    \hline
    Theta range for data collection & 4.749 to 27.508$^\circ$  \\
    Index ranges   & -4$\leq$h$\leq$5 \\ & -5$\leq$k$\leq$5  \\ &  -11$\leq$l$\leq$11  \\
    Reflections collected  &  2035  \\
    Refinement method & Full-matrix least-squares on F$^2$  \\
    Refinement program  & SHELXL-2014 (Sheldrick, 2014)  \\
    Data /restraints /parameters & 114 /0 /12 \\
    Goodness-of-fit on F$^2$  & 1.224 \\
    Final R indices  & R$_1$ = 0.0139\\  &  wR$_2$ = 0.0318 \\
    Weighting scheme  & w=1/[$\sigma^2$(F$_o$$^2$)+0.6368P]  \\
                      & where P=(F$_o$$^2$+2F$_c$$^2$)/3 \\
    Extinction coefficient  & 0.014(3) \\
    Largest diff. peak and hole  &  0.655 and -0.364 e{\AA}$^{-3}$  \\
    R.M.S. deviation from mean  &  0.122 e{\AA}$^{-3}$ \\
    \hline
    \hline
  \end{tabular}
\label{tab.2}
\end{table}

\begin{table}
\centering \caption{Refined lattice constants for the CaFeAsF single
crystal.}
\begin{tabular}{ll}
    \hline
    \hline
    Chemical formula & CaFeAsF  \\
    Formula weight   & 189.85 g/mol  \\
    Temperature & 296(2) K \\
    Wavelength  &  0.71073 {\AA}  \\
    Crystal system &  tetragonal  \\
    Space group   & P4/nmm  (No. 129)  \\
    Z  &  2  \\
    Unit cell dimensions   &  a = 3.8774(4) {\AA},     $\alpha$ = 90$^\circ$  \\
                           &  b = 3.8774(4) {\AA},     $\beta$ = 90$^\circ$  \\
                           &  c = 8.5855(10) {\AA},    $\gamma$ = 90$^\circ$  \\
    Volume  & 129.076(4) {\AA}$^3$\\
    Bond angle ($\delta_{As-Fe-As}$)   &  107.82(6)$^\circ$ $\times$2  \\
                 &  110.30(4)$^\circ$ $\times$4  \\
    Anion height  &  $h_{As}$ = 1.413 {\AA} \\
    Density (calculated)  &  4.885  g/cm$^3$  \\
    Absorption coefficient  & 20.223 mm$^{-1}$ \\
    F(000)   & 176  \\
    \hline
    \hline
  \end{tabular}
\label{tab.3}
\end{table}

\begin{table}
\centering \caption{Atomic coordinates and equivalent isotropic
atomic displacement parameters (\AA$^2$) for CaFeAsF.}
\begin{tabular}{lllll}
    \hline
    \hline
    Atom &  x  &  y  &  z  & U(eq)  \\
    \hline
    As  & 1/4  & 1/4  & 0.16461(8) & 0.0067(3) \\
    Fe  & 3/4  & 1/4  & 0  &  0.0069(3) \\
    Ca  & 3/4  & 3/4  & 0.34801(16)  &  0.0080(4) \\
    F   & 3/4  & 1/4  & 1/2 &  0.0092(8)\\
    \hline
    \hline
  \end{tabular}
\label{tab.4}
\end{table}

We used the high-resolution single-crystal x-ray diffraction to
study the structural details of our sample. The diffraction data
were collected at room temperature by the $\omega$- and
$\varphi$-scan method. The crystal structure was solved by
SHELXS-2014 and refined by SHELX-2014.~\cite{SHELXS-2014} The
parameters for the data collection and structure refinement are
listed in Table II. The values of R$_1$ and wR$_2$ are much smaller
than the previously reported polycrystalline
results,~\cite{CaFeAsF-Co} and also small compared to the Na-doped
single crystalline system,~\cite{Na-doped} indicating the
high-quality of our sample and the reliability of our refinements.
As shown in Table III, the final cell constants are determined to be
$a$ =  $b$ = 3.8774(4) {\AA}, c = 8.5855(10) {\AA}. It is clear that
the c-axis lattice constants are very close to that obtained from
the data in Fig. 2. In addition, the a- and c-axis lattice constants
determined from our experiment are consistent with the
polycrystalline samples reported
previously.~\cite{CaFeAsF-Co,CaFeAsF-Nd} Compared to the Na-doped
single crystalline samples, the a-axis lattice constant is similar
while the c-axis constant is clearly smaller.~\cite{Na-doped} The
anion (As) height relative to Fe layer is a bit larger than the
optimal value (1.38 {\AA}) for the highest $T_c$ in
FeSCs.~\cite{As-height} The atomic coordinates from the refinement
are shown in Table IV, which also confirm the structure obtained
earlier on the basis of powder data, with a difference of about
0.16\% for the c-axis position of As element.~\cite{CaFeAsF-Co} The
bond angles $\delta_{As-Fe-As}$ deviate a bit from the optimal value
of about 109.47$^\circ$.

\begin{figure}
\includegraphics[width=8cm]{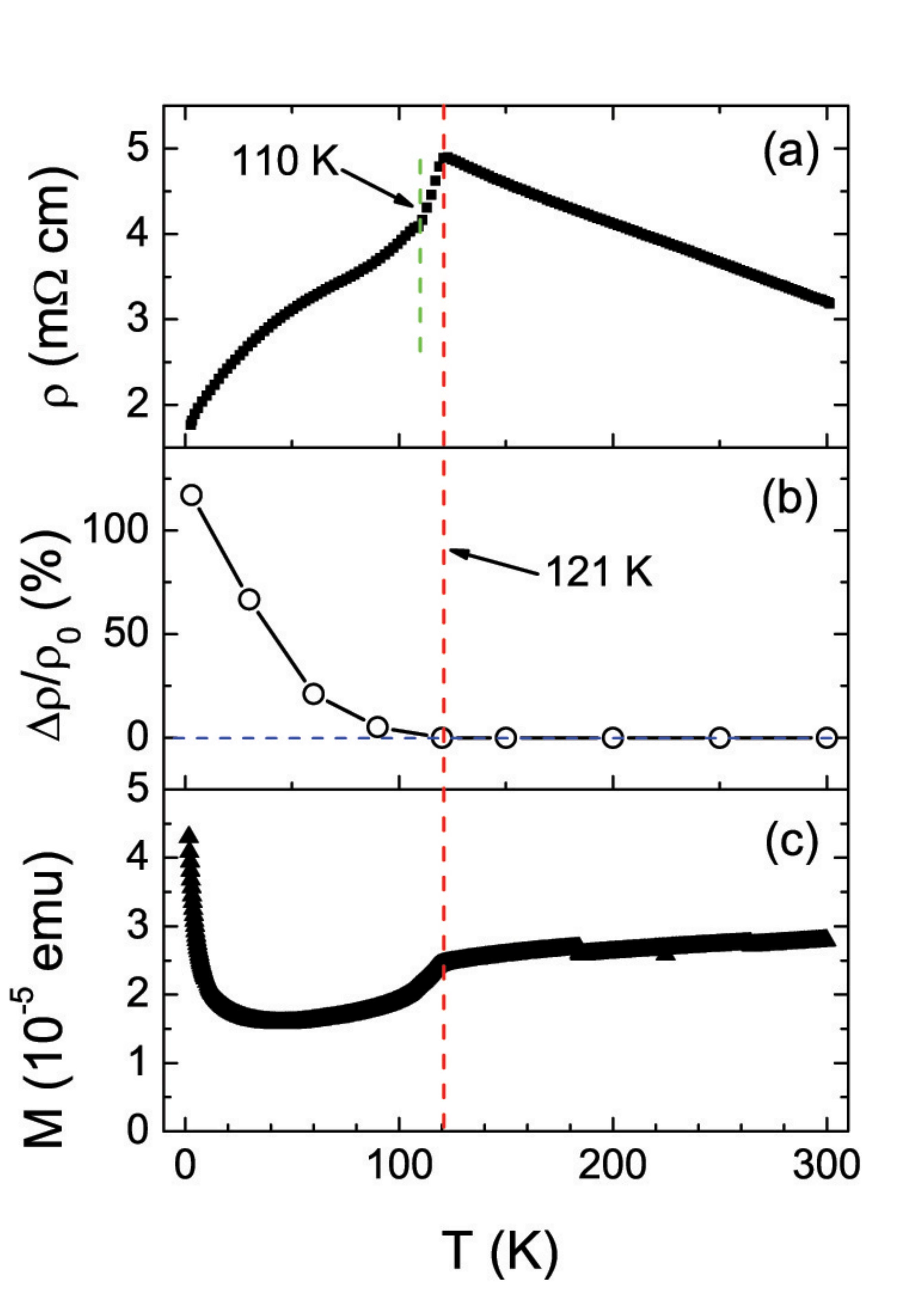}
\caption {(color online) Temperature dependence of the in-plane
resistivity (a), magnetoresistance (b), and the magnetic
susceptibility (c). The field of 1 T was applied along the c-axis of
the crystal during the magnetic susceptibility measurement. The MR
data were collected under the field of 9 T. The dashed lines are
guides for eyes.} \label{fig3}
\end{figure}

The resistivity, MR, and magnetic susceptibility change the
variation tendency at the same temperature 121 K on the temperature
dependent curves, as revealed in Figs. 3(a), (b), and (c). This
seems to be a common feature in most of the FeSCs, associated with
the structure and the spin-density-wave (SDW)-type antiferromagnetic
transition. This transition temperature is a little higher than the
polycrystalline results (118-120 K).~\cite{CaFeAsF-Nd,CaFeAsF-Co}
Temperature dependence of resistivity are shown in Fig. 3(a). Above
121 K, the resistivity increases almost linearly with the decrease
of temperature. We note that it is rather conflicting about this
behavior, among different repots based on polycrystalline
samples.~\cite{CaFeAsF-Co,CaFeAsF-Nd,SrFeAsF1,SrFeAsF2} Only one
result reported on SrFeAsF by Tegel et al. shows similar tendency,
compared with our data.~\cite{SrFeAsF2} We argue that the data from
single-crystal samples reveals the intrinsic properties since the
scattering processes are not affected by the grain boundaries.
Moreover, the transition at 121 K is sharper than the results from
polycrystalline samples. Below that temperature, the resistivity
decreases with cooling and a clear kink can be observed at about 110
K, as indicated by the green dashed line. These two characteristic
temperatures with the interval of 11 K are reminiscent of the
reported $T_{str}$ = 150 K and $T_N$ = 138 K in the another 1111
phase LnFeAsO (Ln = La, Ce),~\cite{LaFeAsO,CeFeAsO} where $T_{str}$
and $T_N$ are the transition temperatures from tetragonal to
orthogonal structure and that from paramagnetic to SDW-type
antiferromagnetic phase, respectively. We note that such two
distinct transition temperatures have also been detected from the
resistivity data in the Co-doped BaFe$_2$As$_2$
system.~\cite{Fisher} So it is very likely that these two transition
temperatures are $T_{str}$ = 121 K and $T_N$ = 110 K for the present
CaFeAsF system.

In is paper, MR is expressed as $\Delta\rho/\rho_0=[\rho(9
\mathrm{T})-\rho_0]/\rho_0$, where $\rho$(9T) and $\rho_0$ are the
resistivity under the field 9 T and zero field, respectively. In
Fig. 3(b), we show temperature dependence of MR. The magnitude of MR
decreases with the increase of the temperature monotonously until
$T_{str}$ and vanishes above this temperature. These observations
suggest that the MR in this system is associated with the magnetic
and electronic structures, which are affected by the structure and
the SDW-type antiferromagnetic transitions remarkably. The
transition on the $M-T$ curve shows the feature of an
antiferromagnetic transition, as shown in Fig. 3(c). In the high
temperature non-magnetic normal state, a
linear-temperature-dependent behavior can be observed. This is a
non-Curie-Weiss-like paramagnetic behavior and cannot be understood
within a simple mean-field picture. This behavior should be very
important to understand the mechanism of high-$T_c$
superconductivity because it was also observed in undoped and highly
underdoped cuprates.~\cite{cuprates} In the pnictide compounds, this
feature was interpreted by the antiferromagnetic fluctuations with
the local SDW correlations.~\cite{TXiang}

\section{Conclusions}
In summary, high-quality and sizable single crystals of
CaFeAsF were grown successfully by the self-flux method. The
single-crystal x-ray diffraction measurements were carried out and
the structure details were refined based on the data. The
resistivity, magnetoresistance, and magnetic susceptibility show
clear different behaviors below and above 121 K. The critical
temperatures of the structure and antiferromagnetic transition were
determined to be $T_{str}$ = 121 K and $T_N$ = 110 K, respectively.
Our results supply a platform to study the intrinsic properties of
the 1111 phase of FeSCs.

\begin{acknowledgments}
This work is supported by the National Natural Science Foundation of
China (No. 11204338), the ``Strategic Priority Research Program (B)"
of the Chinese Academy of Sciences (No. XDB04040300, XDB04040200 and
XDB04030000) and Youth Innovation Promotion Association of the
Chinese Academy of Sciences (No. 2015187).
\end{acknowledgments}

\end{document}